\documentclass[twocolumn]{aastex631}
\usepackage{natbib}
\usepackage[T1]{fontenc}

\usepackage{graphicx}	
\usepackage{amsmath}	
\usepackage{amssymb}	
\usepackage{comment}
\usepackage{cancel}
\usepackage{float}
\usepackage{tikz}
\usepackage{newtxtext,newtxmath}

\newcommand{\xmm}{{\it XMM-Newton}}
\newcommand{\nustar}{{\it NuSTAR}}
\newcommand{\jetcaf}{{\fontfamily{qcr}\selectfont JeTCAF}}
\newcommand{\tcaf}{{\fontfamily{qcr}\selectfont TCAF}}

\newcommand{\plaw}{{\fontfamily{qcr}\selectfont PL}}
\newcommand{\tbabs}{{\fontfamily{qcr}\selectfont TBABS}}

\begin{document}

\title{Broadband X-ray spectral analysis of the ULX NGC\,1313\,X-1 using \jetcaf:\,Origin of the ULX bubble}

\author[0000-0002-4533-3170]{Biswaraj Palit}
\affiliation{Indian Institute of Astrophysics, II Block, Koramangala, Bangalore 560034, India}

\affiliation{Nicolaus Copernicus Astronomical Center, Polish Academy of Sciences,Bartycka 18, 00716, Warsaw, Poland}

\author[0000-0003-0793-6066]{Santanu Mondal}
\affiliation{Indian Institute of Astrophysics, II Block, Koramangala, Bangalore 560034, India}
\affiliation{Equals first author}

\correspondingauthor{Santanu Mondal}
\email{santanuicsp@gmail.com}



\begin{abstract}

NGC\,1313\,X-1 is a mysterious Ultra-luminous X-ray (ULX) source whose X-ray powering mechanism and a bubble-like structure surrounding the source are topics of intense study. Here, we perform the X-ray spectroscopic study of the source using a joint {\it XMM-Newton} and {\it NuSTAR} observations taken during 2012 $-$ 2017. The combined spectra cover the energy band 0.3 $-$ 20 keV. We use the accretion-ejection-based \jetcaf\, model for spectral analysis. The model fitted disc mass accretion rate varies from \textcolor{black}{4.6 to 9.6} $\dot M_{\rm Edd}$ and the halo mass accretion rate varies from \textcolor{black}{4.0 to 6.1} $\dot M_{\rm Edd}$ with a dynamic Comptonizing corona of average size of \textcolor{black}{$\sim 15$} $r_g$. \textcolor{black}{The data fitting is carried out for different black hole (BH) mass values. The goodness of the fit and uncertainties in model parameters improve while using higher BH mass with most probable mass of the compact object to be $133\pm33$ M$_\odot$}. We have estimated the mass outflow rate, its velocity and power, and the age of the inflated bubble surrounding the source. Our estimated bubble morphology is in accord with the observed optical bubble and winds found through high-resolution X-ray spectroscopy, suggesting that the bubble expanded by the outflows originating from the central source. Finally, we conclude that the super-Eddington accretion onto a nearly intermediate mass BH may power a ULX when the accretion efficiency is low, though their efficiency increases when jet/outflow is taken into account, in agreement with numerical simulations in the literature.

\end{abstract}



\section{Introduction} \label{sec:intro}

Ultra-luminous X-ray sources (ULXs) are point-like sources with isotropic luminosities exceeding a value of $10^{39}$ erg s$^{-1}$. To date, a few hundred of ULXs are known \citep{SwartzEtal2004ApJS..154..519S,WaltonEtal2011MNRAS.416.1844W}. A large number of ULXs are located in star-forming galaxies and associated with young stellar population \citep[][]{FabbianoEtal2001ApJ...554.1035F,SwartzEtal2009ApJ...703..159S,PoutanenEtal2013MNRAS.432..506P}. However, their powering mechanism is not yet well-understood. So far, different scenarios have been proposed to explain various observational features including the luminosity of ULXs.

First of them involves, super-Eddington accretion (with or without beaming) onto a stellar mass black holes \citep[StMBH;][]{GilfanovEtal2004NuPhS.132..369G,poutenan2007MNRAS.377.1187P,King2009MNRAS.393L..41K}. A key feature predicted by the theory and simulation \citep[][and references therein]{poutenan2007MNRAS.377.1187P,TakeuchiEtal2013PASJ...65...88T,KobayashiEtal2018PASJ...70...22K} for this type of  accretion is the presence of strong optically thick wind, which covers the inner region of the disc and collimates the radiation and also from observations \citep[][and references therein]{middle2015MNRAS.454.3134M}.
While these models give clues to understanding the super-Eddington accretion regime to some extent, many questions about the super-Eddington regime and its connection with ULXs remain open. For instance, 1) what is the degree to which emission is beamed \citep[e.g.][and references therein]{King2001ApJ...552L.109K,JiangEtal2014ApJ...796..106J,MushtukovEtal2021MNRAS.501.2424M}? 2) how much fraction of energy is carried out as outflows? 3) what are the mechanical and radiative feedback induced by ULXs? and 4) what is the exact accretion flow geometry allowing these objects to reach such high luminosity? Conversely, if the StMBH has a highly magnetized accretion disc, then even sub-Eddington accretion can power some ULXs \citep{TusharBani2019MNRAS.482L..24M}.

The second scenario is sub-Eddington accretion onto so-called intermediate mass black holes \citep[IMBH;][and references therein]{colbert1999ApJ...519...89C,miller2003ApJ...585L..37M}. This accretion regime is typical for Galactic Black Hole Binaries (GBHBs), therefore they could show similar properties in accretion \citep{KaaretEtAl.2001MNRAS.321L..29K,miller2003ApJ...585L..37M}. However, these IMBHs may accrete in super-Eddington regime and power some ULXs \citep{MondalEtal2022JAA}. For instance, by studying the Chandra observations of Antennae galaxy, \citet[][]{King2001ApJ...552L.109K} proposed that under certain conditions on stellar companion and binary orbit would allow the possibility that individual ULXs may harbor extremely massive black holes (MBHs), the growth of massive BHs can also be through rapid mass accretion in $\sim$ 100 M$_\odot $ BHs \citep[][for a review]{jenny2020ARA&A..58..257G}, after the death of the earliest known Pop-III stars. 

However, while it is generally accepted the above two scenarios, the discovery of X-ray pulsations in one ULX \citep{Bachetti2014Natur.514..202B} showed that neutron star (NSs) can also attain super-Eddington luminosities. Followed by the discovery, a few more pulsating ULXs \citep[PULXs;][]{FurstEtal2016ApJ...831L..14F,isreal2017Sci...355..817I,CarpanoEtal2018MNRAS.476L..45C,Satya2019MNRAS.488L..35S,RodriguezCastilloEtal2020ApJ...895...60R} and the possible confirmation of another NSULX through the detection of a cyclotron resonance by strong magnetic field \citep{BrightmanEtal2018NatAs...2..312B} have been identified. These discoveries and findings suggest that NSULX may dominate the ULX population. Yet, there is still some debate on the underlying powering mechanism for such extreme luminosities.  

NGC\,1313\,X-1 (hereafter ULX-1) is located in the starburst galaxy NGC\,1313 at a distance of 4.13 Mpc \citep{mendez2002AJ....124..213M}. The galaxy also hosts other prominent luminous sources, however, the ULX-1 can be well isolated from other sources, suffers less from background contamination, and is also in proximity to the Earth (z $\sim$ 0.00157). This provides a unique opportunity to obtain observationally rich information. 
ULX$-$1 has been extensively studied in the spectro-temporal domain in the literature. \citet{feng2006ApJ...650L..75F} studied the spectral evolution of both ULX sources (X-1 and X-2) using simple powerlaw (\plaw)\, continuum and multi-color disc (MCD) models within the energy range of {\it XMM-Newton} before ULX-2 was identified as a likely pulsar \citep{Satya2019MNRAS.488L..35S}. Recently, 
\citet{walton2020MNRAS.494.6012W} analysed combined multi-instrument {\it XMM-Newton+Chandra+NuSTAR} spectra of ULX-1 to study the broadband spectral variability using three component disc model. 
A variability analysis was conducted between different energy bands to understand the causal connection between different emission regions \citep{Erin2020MNRAS.491.5172K} in the accretion disc.
\citet{Gald10.1111/j.1365-2966.2009.15123.x} reported a spectral cutoff at $\sim$ 5 keV. For the first time, \citet{bachetti2013ApJ...778..163B}, studied ULX-1 using joint {\it XMM-Newton} and {\it NuSTAR} data and suggested a spectral break above 10 keV, where the BH is accreting in near Eddington rate. Along with the continuum spectral variability, emission and absorption lines have been observed too for the ULX$-$1 \citep{pinto2016Natur.533...64P,walton2016ApJ...826L..26W,Pinto2020MNRAS.492.4646P}. \citet[][]{Erin2020MNRAS.491.5172K} took an attempt to explain timing properties as originating from beamed relativistic outflows. Very recently, a shock ionized bubble has been identified around the ULX-1 using MUSE spectroscopic studies \citep{gurpide2022arXiv220109333G}, which suggests the presence of outflows from the ULX-1. Similar bubble structure was reported earlier by \citet{Pakull2008AIPC.1010..303P} in other ULX systems.

Several studies in the literature put light on the mass estimation of the central compact object in ULX$-$1. 
Those findings reported two possibilities of the mass of the BH, one in the StMBH to the higher end of the StMBH range \citep{miller2003ApJ...585L..37M,Soria2007Ap&SS.311..213S,bachetti2013ApJ...778..163B}, while the other in the IMBH range \citep{miller2003ApJ...585L..37M,fabian2004MNRAS.355..359F,wang2004ApJ...609..113W,dewangan2010MNRAS.407..291D,jang2018MNRAS.473..136J,KobayashiEtal2019MNRAS.489..366K}. The quasi-periodic oscillation study suggested mass in the IMBH range \citep{PashamEtAl2015ApJ,HuangEtAl2019CoSka}. Overall, we see that the mass of the ULX-1 is reported over a very large range, from as low as 11 M$_\odot$ to as high as 7000 M$_\odot$. Therefore, the type of the central compact object is not known to date. Hence, to understand the importance of these differences in opinions, an extensive study on the central object is required. 

Here, we highlight some of the observed signatures and evidence, that direct us to consider ULX-1 as a likely BH accretor: (1) The color-color diagram in \citet[][]{pintore2017ApJ...836..113P} shows that ULX-1 is situated at the centre of the plot while extending towards softer ratios. Moreover, they suggested that ULX-1 might not host a NS accretor. This is firstly supported by the non-detection of pulsations till date, (2) \citet{walton2020MNRAS.494.6012W} carried out extensive pulsation search with both \xmm\, and \nustar\, data of ULX-1, however, did not detect any signal above 3-$\sigma$ confidence level. A similar conclusion was drawn by \citet{DoroshenkoEtal2015A&A...579A..22D}. The non-detection of pulsation could be due to the limited statistics, low pulse-period, and variable pulsation, which could be improved with additional observation. It can also be possible that the signal is faded by scattering from the wind, (3) According to \citet[][]{gurpide2021A&A...649A.104G}, a BH accretor can swallow any excess radiation in its vicinity, in the process of stabilising the outflowing radiation. Thus, the absence of large variability in hard energy ranges disfavors the presence of NS accretor, and (4) A dipole field strength of $\le 10^{10}$~G calculated for ULX-1 considering propellar state transitions \citep{MiddletonEtal2022MNRAS.tmp.3156M} is quite low compared to some PULXs.Therefore, we carry out the rest of the analysis of broadband X-ray of ULX-1 considering it a BH candidate.

To understand the rich accretion behavior, several authors have undertaken combined disc-corona models in their study. These models successfully fit the spectra and extract the corona properties. However, most of them are solely radiative transfer mechanism based and disregard the physical origin of the corona and change in its properties (optical depth, size, temperature, etc.). Therefore, it motivates us to use a model which self-consistently considers both disc, dynamic corona, and mass outflow in a single picture.  

\begin{figure}
    \centering
    \includegraphics[height=6.5cm]{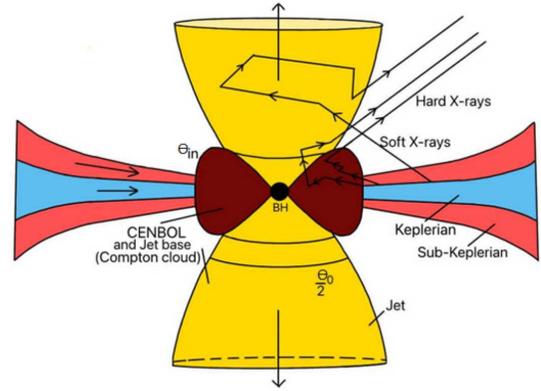}
    \caption{Illustration of the \jetcaf\, model. The blue and red colors show the Keplerian and sub-Keplerian flows respectively. Brown colored region is the inflated, hot CENBOL region. The yellow color shows the ejection of jet. The zig-zag arrows show the scattering of disc photons by different Comptonizing mediums. The figure is adapted from \citep{santanu2021ApJ...920...41M}}
    \label{fig:jetcaf-cartoon}
\end{figure}

According to the Two Component Advective Flow (\tcaf\,) solution \citep{Chakravartti1995ApJ...455..623C}, accretion disc has two components one is a standard, high viscosity, optically thick Keplerian disc and the other one is a hot, low viscosity, optically thin sub-Keplerian disc. The second component moves faster and forms the inner dynamic corona after forming a hydrodynamic shock \citep[][and references therein]{Fukue1987PASJ...39..309F,Chakrabarti1989ApJ...347..365C,MondalChakrabati2013MNRAS.431.2716M}. In the post-shock region (or dynamic corona) which is also known as CENBOL (CENtrifugal BOundary Layer), matter piles up there and soft photons from the Keplerian disc get upscattered to become hard due to inverse Comptonisation. 
This model does not include the effects of jet/mass outflow, which is believed to be originated from the base of the dynamic corona \citep{Chak1999A&A...351..185C}. Very recently, \citet{santanu2021ApJ...920...41M} implemented jet/mass outflow in \tcaf\, (\jetcaf) solution to examine its effect on the emission spectra. The cartoon diagram of the model is shown in \autoref{fig:jetcaf-cartoon}. 

The \jetcaf\, model has six parameters, including BH mass, if the mass of the central compact object is not known. These parameters are namely (1) mass of the BH ($M_{\rm BH}$), (2) Keplerian mass accretion rate ($\dot m_d$), (3) sub-Keplerian mass accretion rate ($\dot m_h$), (4) size of the dynamic corona or the location of the shock ($X_s$), (5) shock compression ratio (R=post-shock density/pre-shock density), and (6) jet/mass outflow collimation factor ($f_{\rm col}$=solid angle subtended by the outflow/inflow). Therefore, one can estimate the outflowing opening angle using this parameter. Based on the opening angle it can be inferred whether the outflow is collimated or not.

In this paper, we aim to analyze the joint {\it XMM-Newton} and {\it NuSTAR} data and fit them using \jetcaf\, model to understand the accretion-ejection properties around the ULX-1. The recent discovery of optical bubbles also motivated us to estimate the jet/mass outflow properties in this system using the \jetcaf\, model. In addition, as the mass of the central BH is still under debate, our study also puts some light on the possible mass estimation of the central BH.
In the next section, we discuss the observation and data analysis procedures. In section 3, the model fitted results along with the estimation of different accretion-ejection flow quantities are discussed. We also discuss some of the limitations of the model, X-ray data analysis of ULXs, and the model dependence of the results. Finally, we draw our brief conclusion.

\section{Observation and Data Reduction} 

We used all available joint \xmm\, and \nustar\, \citep{Harrisonetal2013} observations of ULX-1 during 2012 to 2017. The log of observations is given in \autoref{tab:obslog}.

The \xmm\,data are reprocessed using Science Analysis System (SAS) version 19.1.0 and followed the standard procedures given in the SAS data analysis threads\footnote{\url{https://www.cosmos.esa.int/web/xmm-newton/sas-threads}}. 
As a first step, {\it epproc} routine was executed to generate the calibrated and concatenated event lists. The filtering of the data from background particles was done by selecting a source free region in the neighbourhood of ULX-1.
Then we viewed the filtered image using {\sc saods9} software to select the source and background regions. 
An extraction region of radius 30" circling the source ULX-1 as well as a nearby background region free of any sources was taken into account. The source and background spectra were produced by restricting patterns to singles and doubles, followed by the ``rmf'' and ``arf'' file generation using the standard {\sc rmfgen} and {\sc arfgen} task. We extracted the source and background spectra using the {\sc evselect} routine. Finally, we rebinned the spectra using {\sc specgroup} task to have a minimum of 35 counts in each bin. For the analysis of each epoch of observation, we used the data of {\it XMM-Newton} in the energy range of 0.3$-$10\,keV, above 10 keV, data is noisy. 
The \nustar\, data were extracted using the standard 
{\sc NUSTARDAS}
\footnote{\url{https://heasarc.gsfc.nasa.gov/docs/nustar/analysis/}} software. We 
ran {\sc nupipeline} task to produce cleaned event lists and 
{\sc nuproducts} to generate the spectra. The data were grouped by {\sc grppha} command, with a 
minimum of 35 counts in each bin. For the analysis of each epoch of observation, we used the data of {\it NuSTAR} in the energy range of 3$-$20\,keV. The data is noisy above 20 keV. 

We used {\sc XSPEC}\footnote{\url{https://heasarc.gsfc.nasa.gov/xanadu/xspec/}} \citep{Arnaud1996} version 12.11.0 for spectral analysis. 
Each epoch of the joint observation was fitted using \jetcaf\, model in the total energy range 0.3$-$20 keV along with a single neutral hydrogen absorption column (using \tbabs\, model) component. \textcolor{black}{We used \textsc{wilms} abundances \citep{Wilmsetal2000} and cross-section of \citet{Verner1996} in our analysis.}
We used chi-square statistics for the goodness of the fitting.

\begin{table}
    \centering
     \caption{Observation log of joint {\it XMM-Newton} and {\it NuSTAR} data.}
    \begin{tabular}{c c c c C}
    \hline
    Epoch & ObsID            & ObsID        & Date & \textcolor{black}{MJD}\\
          & {\it XMM-Newton} & {\it NuSTAR} &  & \\
    \hline
    A1 & 0803990601 & 30302016010 & 2017-12-09 &58096\\
    A2 & 0803990101 & 30302016002 & 2017-06-14 &57918\\
    A3 & 0794580601 & 90201050002 & 2017-03-29 &57841\\
    A4 & 0742590301 & 80001032002 & 2014-07-05 &56843\\    
    A5 & 0693850501 & 30002035002 & 2012-12-16 &56277\\
    \hline
   \end{tabular}
    \label{tab:obslog}
  \end{table}

\section{Results and Discussions} 

\subsection{Spectral Fitting} \label{sec:SpecFit}
\begin{figure*}
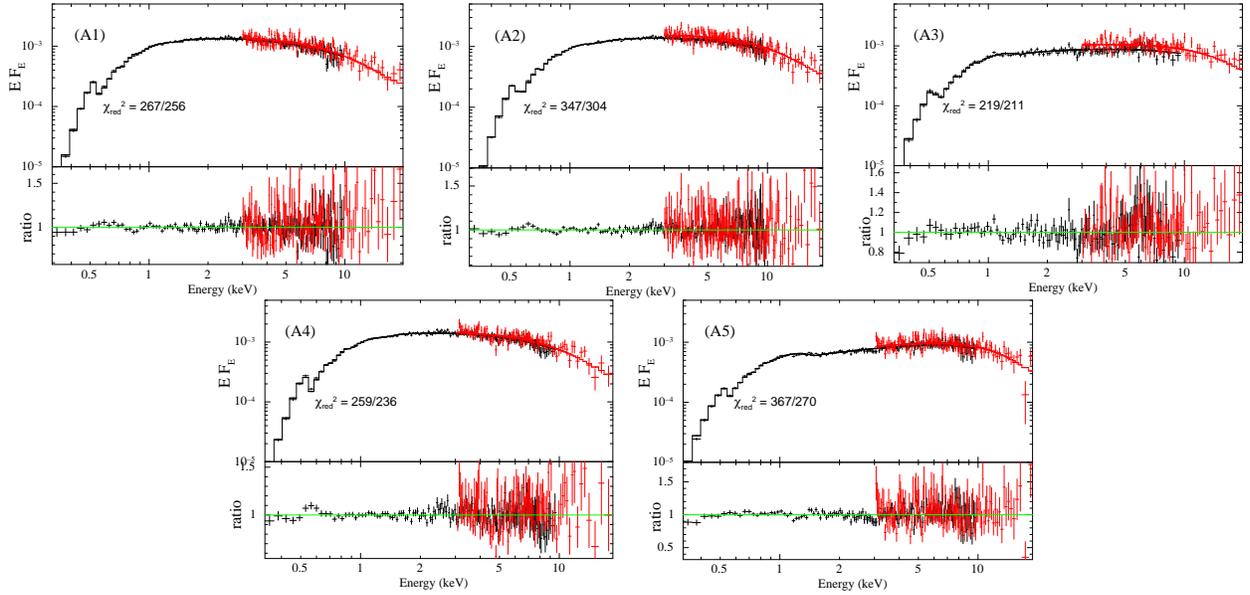

    \centering{
    \includegraphics[height=5.5cm,angle=270]{Figures/r1-A1-jetcaf-v2.eps}
    \includegraphics[height=5.5cm,angle=270]{Figures/r1-A2-jetcaf-v2.eps}
    \includegraphics[height=5.5cm,angle=270]{Figures/r1-A3-jetcaf-v2.eps}
    \includegraphics[height=5.5cm,angle=270]{Figures/r1-A4-jetcaf-v2.eps}
    \includegraphics[height=5.5cm,angle=270]{Figures/r1-A5-jetcaf-v2.eps}}
    \caption{Spectral fitting for all five epochs with \jetcaf\, model are shown. The upper segment is the spectra while the bottom segment is the ratio. The black-colored data corresponds to {\it XMM-Newton} EPIC-PN while red-colored data corresponds to {\it NuSTAR}. \textcolor{black}{The slight misalignment between \xmm\, and \nustar\, data is due to residual cross-calibration and possibly the non-perfect simultaneity of the observations.}}
    \label{fig:jetcafspec}
\end{figure*}

\textcolor{black}{All epochs of data in the range from $0.3-20$ keV are fitted using \jetcaf\,model considering the mass of the BH as a free parameter (hereafter model M1) and keeping its value fixed to 10, 30, and 100 M$_\odot$, which we denote as model M2, M3, and M4 respectively. All other model parameters are left free to vary during fitting including the model normalization (``Norm''). We fixed the constant for EPIC detectors of the \xmm\, satellite to 1 and left it free for the \nustar\,to determine a cross-calibration constant. This takes into account residual cross-calibration between \xmm\,and \nustar\, and the possible mismatches due to strictly non-simultaneous observations. The cross-normalization constant between \nustar\, and \xmm\, spectra is obtained between $1.05\pm0.04 -1.20\pm0.07$ for all epochs using model M1. Other models (M2-M4) also showed a similar range of values. \autoref{fig:jetcafspec} shows the M1 model fitted spectra to the data. The spectra in the epochs A2 and A4 are looking alike, while A3 and A5 are similar, however, A1 appears to be in between those two shapes. Therefore it can be possible that during those epochs, the source passed through the same spectral states. We have discussed this later. The best fitted M1 model fitted parameters are given in \autoref{table:JeTCAFResults}.} 

\begin{table*}
\centering
\caption{\label{table:JeTCAFResults} \textcolor{black}{The broadband spectral parameters of NGC\,1313\,X-1 when fitted with \jetcaf\,model. The M$_{\rm BH}$, $\dot m_d$, $\dot m_h$, $X_s$, $R$, $f_{\rm col}$ and Norm are the mass of the black hole, disk and halo mass accretion rates, location of the shock or size of the corona, shock compression ratio, jet collimation factor, and model normalization respectively.} Here, N$_H$ is the hydrogen column density along the LOS.}
\resizebox{\textwidth}{!}{\begin{tabular}{cccccccccc}
\hline
Obs.Ids. &M$_{\rm BH}$ &$\dot m_{\rm d}$ & $\dot m_{\rm h}$ & $X_{\rm s}$ & R &$f_{\rm col}$&Norm&N$_H$&$\chi^2/dof$ \\
         & $M_\odot$&$\dot M_{\rm Edd}$&$\dot M_{\rm Edd}$&$r_{\rm g}$& & & &$\times 10^{22}$ cm$^{-2}$& \\
\hline
A1 &$163.2 \pm 21.2$        &$4.61 \pm 0.44$&$3.96 \pm 0.26$&$14.9 \pm 2.1$&$5.1 \pm 0.3$&$0.77 \pm 0.06$&$0.33\pm0.08$&$0.28 \pm 0.02$&267/256 \\
A2 &$100.1 \pm  7.1$        &$6.11 \pm 0.71$&$4.22 \pm 0.26$&$16.9 \pm 1.1$&$4.4 \pm 0.2$&$0.69 \pm 0.05$&$0.57\pm0.19$&$0.27 \pm 0.02$&347/304 \\
A3 &$128.5 \pm 18.5$        &$9.55 \pm 1.86$&$5.21 \pm 0.79$&$15.0 \pm 2.6$&$3.2 \pm 0.8$&$0.60 \pm 0.06$&$0.25\pm0.06$&$0.22 \pm 0.04$&219/211 \\
A4 &$166.3 \pm 26.4$        &$4.58 \pm 0.51$&$3.95 \pm 0.28$&$14.7 \pm 2.3$&$5.2 \pm 0.3$&$0.79 \pm 0.07$&$0.32\pm0.09$&$0.28 \pm 0.02$&259/236 \\
A5 &$106.0 \pm  8.1$        &$8.90 \pm 0.86$&$6.05 \pm 0.30$&$13.3 \pm 2.1$&$3.7 \pm 0.4$&$0.61 \pm 0.06$&$0.18\pm0.01$&$0.17 \pm 0.01$&367/270 \\
\hline
\end{tabular} }
\end{table*}

\begin{table*}
\centering
\caption{\label{table:ResMfixed} \textcolor{black}{The broadband spectral parameters of NGC\,1313\,X-1 fitted with \jetcaf\,model keeping the M$_{\rm BH}$ parameter fixed (denoted by $^f$) to 10, 30, and 100 M$_\odot$.}}
\resizebox{\textwidth}{!}{\begin{tabular}{cccccccccc}
\hline
Obs.Ids. &$M_{\rm BH}$ &$\dot m_{\rm d}$ & $\dot m_{\rm h}$ & $X_{\rm s}$ & R &$f_{\rm col}$&Norm&$N_H$&$\chi^2/dof$ \\
         & $M_\odot$&$\dot M_{\rm Edd}$&$\dot M_{\rm Edd}$&$r_{\rm g}$& & & &$\times 10^{22}$ cm$^{-2}$& \\
\hline
A1 &        &$5.18 \pm 1.36$&$4.11 \pm 0.75$&$16.9 \pm 1.1$&$5.2 \pm 0.6$&$0.77 \pm 0.18$&$4.38\pm0.76$&$0.16 \pm 0.01$&300/257 \\
A2 &        &$6.57 \pm 0.77$&$3.89 \pm 0.35$&$23.4 \pm 1.8$&$4.5 \pm 0.3$&$0.76 \pm 0.06$&$5.22\pm0.38$&$0.20 \pm 0.01$&354/305 \\
A3 &$10^f$  &$10.01 \pm 1.75$&$3.61 \pm 0.34$&$21.2 \pm 4.8$&$1.9 \pm 0.7$&$0.99 \pm 0.51$&$2.66\pm1.29$&$0.13 \pm 0.02$&307/212 \\
A4 &        &$4.21 \pm 1.57$&$3.28 \pm 0.89$&$18.4 \pm 2.1$&$5.5 \pm 0.8$&$0.99 \pm 0.37$&$3.94\pm1.33$&$0.17 \pm 0.02$&277/237 \\
A5 &        &$4.97 \pm 0.44$&$3.27 \pm 0.20$&$24.4 \pm 2.7$&$4.9 \pm 0.2$&$0.97 \pm 0.10$&$2.05\pm0.18$&$0.12 \pm 0.01$&512/271 \\
\hline
A1 &        &$5.19 \pm 0.94$&$3.85 \pm 0.40$&$17.8 \pm 1.6$&$5.0 \pm 0.4$&$0.79 \pm 0.10$&$1.66\pm0.18$&$0.23 \pm 0.01$&272/257 \\
A2 &        &$6.16 \pm 0.64$&$3.84 \pm 0.30$&$21.1 \pm 3.6$&$4.5 \pm 0.2$&$0.75 \pm 0.05$&$1.86\pm0.23$&$0.24 \pm 0.02$&346/305 \\
A3 &$30^f$  &$10.72 \pm 3.67$&$3.86 \pm 1.87$&$22.3 \pm 6.7$&$2.0 \pm 0.6$&$0.97 \pm 0.30$&$0.92\pm0.59$&$0.18 \pm 0.04$&252/212 \\
A4 &        &$4.01 \pm 0.82$&$3.30 \pm 0.76$&$18.1 \pm 4.2$&$5.5 \pm 0.5$&$0.99 \pm 0.26$&$1.39\pm0.37$&$0.23 \pm 0.03$&266/237 \\
A5 &        &$7.84 \pm 0.66$&$4.53 \pm 0.48$&$22.9 \pm 4.7$&$4.0 \pm 0.5$&$0.73 \pm 0.10$&$0.77\pm0.10$&$0.16 \pm 0.02$&383/271 \\
\hline
A1 &        &$4.94 \pm 0.49$&$3.88 \pm 0.39$&$17.9 \pm 2.9$&$5.0 \pm 0.4$&$0.78 \pm 0.10$&$0.58\pm0.12$&$0.28 \pm 0.02$&267/257 \\
A2 &        &$6.10 \pm 0.71$&$4.22 \pm 0.26$&$16.9 \pm 1.0$&$4.4 \pm 0.2$&$0.69 \pm 0.04$&$0.57\pm0.04$&$0.27 \pm 0.07$&347/305 \\
A3 &$100^f$ &$11.82 \pm 2.54$&$4.54 \pm 1.03$&$21.8 \pm 4.9$&$2.1 \pm 0.4$&$0.84 \pm 0.24$&$0.30\pm0.17$&$0.22 \pm 0.02$&239/212 \\
A4 &        &$4.12 \pm 0.72$&$3.57 \pm 0.34$&$16.4 \pm 2.1$&$5.4 \pm 0.3$&$0.89 \pm 0.11$&$0.49\pm0.10$&$0.27 \pm 0.01$&260/237 \\
A5 &        &$8.91 \pm 0.88$&$6.05 \pm 0.36$&$13.3 \pm 1.4$&$3.7 \pm 0.3$&$0.61 \pm 0.02$&$0.19\pm0.01$&$0.17 \pm 0.02$&369/271 \\
\hline
\end{tabular} }
\end{table*}



\begin{figure}
\hspace{-0.3cm}
\begin{tikzpicture}
\draw (0, 0) node[inner sep=0] {\raisebox{0.5cm}{\includegraphics[height=9cm,trim={0.6cm 0.5cm 1.8cm 2.3cm}, clip]{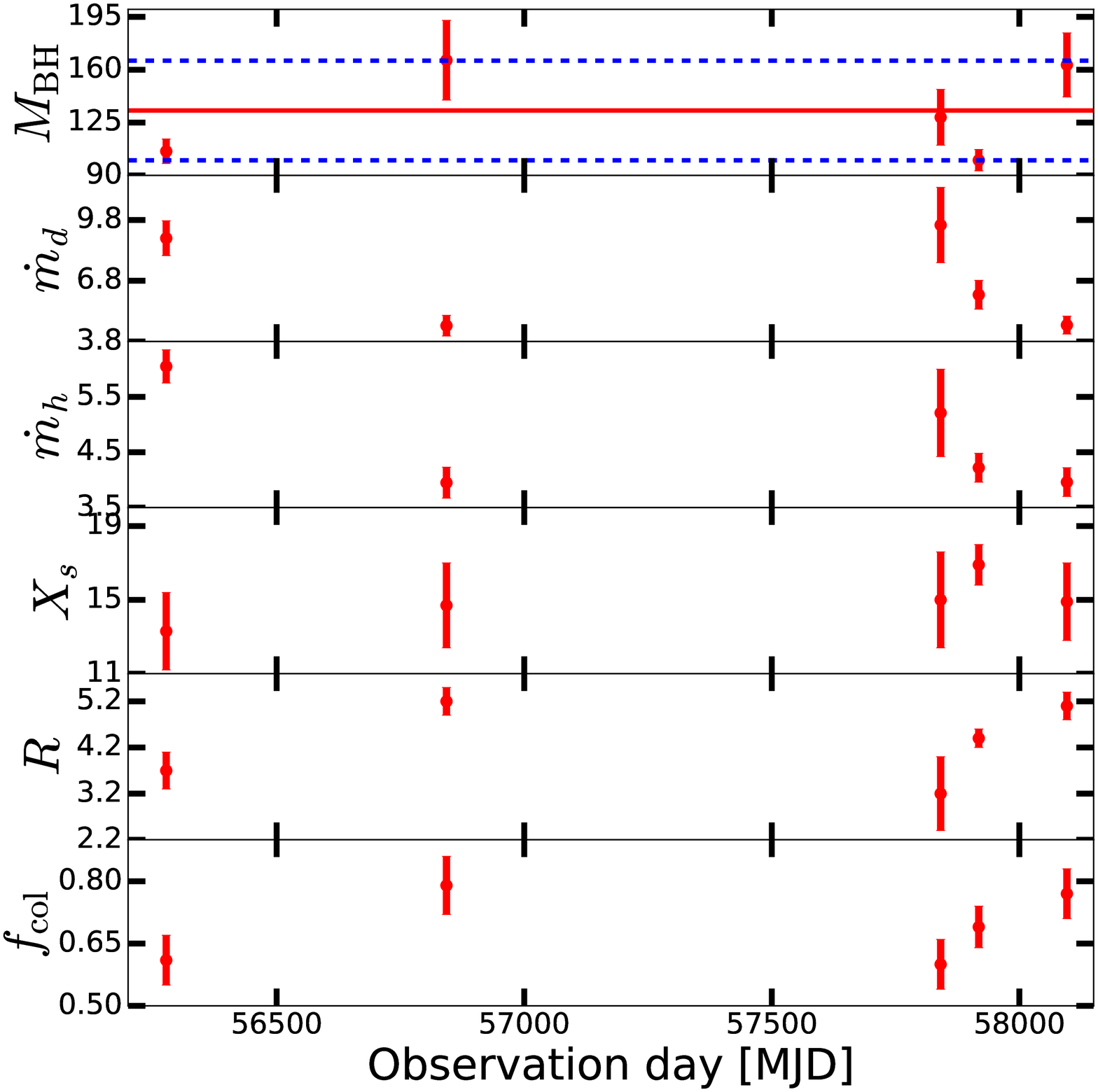}}};
\end{tikzpicture}
\vspace{-1.0cm}
\caption{Best fitted \jetcaf\, model parameters variation with MJD are shown.}   \label{fig:jetcaf}
\end{figure}


\begin{figure}
    \centering{
    \includegraphics[height=10cm,width=8.5cm]{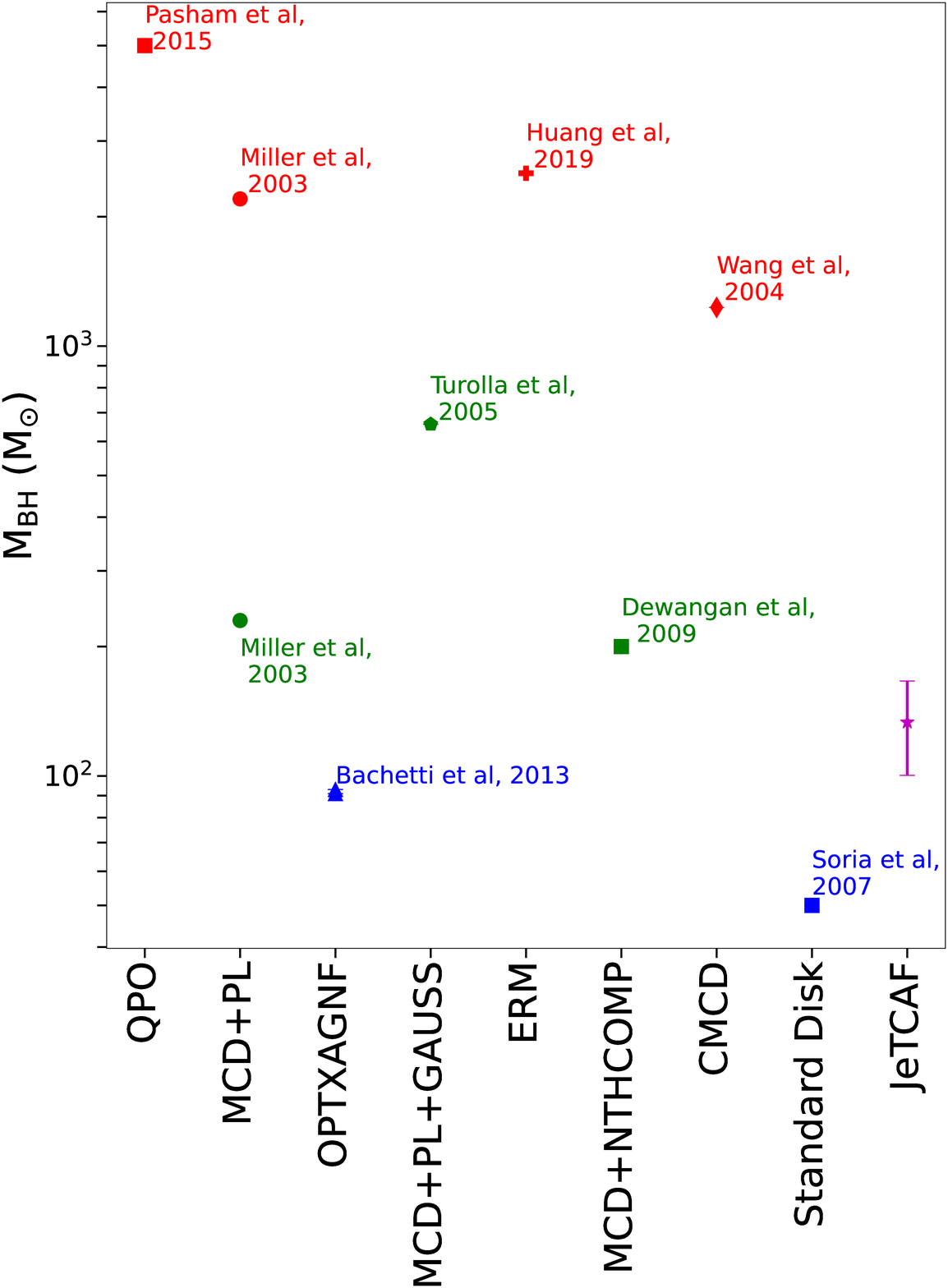}}
    \caption{A comparion of BH mass estimated in this work and other works in the literature. Red, green, and blue colored data points represent mass estimates above 1000M$_\odot$, between 1000 M$_\odot$-100 M$_\odot$ and less than 100 M$_\odot$ respectively. The X-axis indicates the models used to estimate the mass. \textcolor{black}{The magenta point with error bar represents the mass estimated in this work} }
    \label{fig:massplot}
\end{figure}

\textcolor{black}{\autoref{fig:jetcaf} shows the variation of M1 model fitted parameters with MJD. The top to bottom rows indicates the mass of the black hole, mass accretion rates, shock compression ratio, size of the dynamic corona, and the jet/mass outflow collimation factor respectively. The black hole mass obtained from the fit varies in a range between 100-166 M$_\odot$ with an average of $133\pm33$ M$_\odot$, marked by the red solid line with blue dashed lines as uncertainties in mass estimation (in the top panel). The disc mass accretion rate varies in the super-Eddington regime between $\sim$ 4.6 to 9.6 $\dot M_{\rm Edd}$ and the halo accretion rate is also in the super-Eddington regime $\sim$ 4.0 to 6.1 $\dot M_{\rm Edd}$. The size of the dynamic corona/shock location varies between 13 to 17 $r_g$ and the shock compression ratio changes significantly in the range of 3.2-5.2. The $f_{\rm col}$ value is moderately high, fluctuating between 0.6 to 0.8. We kept hydrogen column density (N$_{\rm H}$) parameter free during the fitting and obtained its value of 0.17-0.28, consistent with other works in the literature \citep{walton2020MNRAS.494.6012W}. Overall, it can be seen that the parameters are showing two parallel profiles; during 2012-2014 with 2017. It is likely that the accretion flow behaviour and spectral properties are returned back in 2017 after 3 years. This could be verified if we would have continuous observations of the source. The reduced $\chi^{2}$ ($\chi_r^{2}$) value obtained from the fit is $\sim 1$ for all epochs except the epoch E, where the $\chi_r^{2}$ is $\sim 1.4$. } 

\textcolor{black}{To further check the goodness of the spectral fit and to verify the mass of the BH, we fit the data using model M2-M4. We notice that for the model M2, the fit is relatively poor ($\chi_r^2 \sim 1.2-1.9$) and the uncertainties to the parameters are high. The fit has improved for the model M3 with $\chi_r^2 \sim 1.1-1.4$. A similar goodness of the fit is obtained while using model M4, however, the uncertainties in model parameters have improved while increasing the M$_{\rm BH}$ parameter value. Furthermore, the parameters in models M1 and M4 are similar within error bar, showing a convergence in spectral fitting parameters, thereby, the \jetcaf\, parameters seem robust. All model parameter values, goodness of the fit and the uncertainties in model parameters are given in \autoref{table:ResMfixed}. Therefore, based on the mass dependence study and the robustness of the parameters, it can be said that the ULX1 is harbouring a black hole of mass at the lower end (nearly) of the intermediate mass range. However, a longterm and daily basis spectro-timing study may give a robust estimation with smaller uncertainties.} 

\textcolor{black}{In \autoref{fig:massplot}, we show the comparison of BH mass obtained from the model fit in this work (the magenta line with error bar) with the estimations using different models in the literature. We note that, as the luminosity is a product of accretion efficiency ($\eta_{\rm acc}$), M$_{\rm BH}$, and the mass accretion rate, the overall luminosity may scale up/down depending on the increasing/decreasing individual parameter values. Thereby, it is likely to have a degeneracy in results, which might be the scenario in M$_{\rm BH}$ estimation using phenomenological scaling relations. On the contrary, the shape of an observed spectrum is distinctive, therefore direct fitting of the spectrum using M$_{\rm BH}$, and accretion rate parameters can minimize the degeneracy to some limit which is the case in \jetcaf\,model. Here, we are simultaneously solving a series of equations and finally getting the spectrum. A noticeable change in accretion rate changes the spectral shape, that may not be able to fit the observed spectrum with good statistics. Also, comparing the parameter values in \autoref{table:JeTCAFResults} and \ref{table:ResMfixed} show that they are converging for higher M$_{\rm BH}$ values and lower uncertainties. This may infer that the estimated model parameters are minimally degenerate.} 

\textcolor{black}{Considering the model fitted (from \autoref{table:JeTCAFResults}) $M_{\rm BH}$ and total mass inflow rate ($\dot m_{\rm in}$) as $\dot m_d + R \dot m_h$ \citep{Mondaletal2014Apss}, the accretion luminosity can be estimated to be 3.2-5.4 $\times 10^{41}$ erg s$^{-1}$. However, the observed luminosity ($L_X$) obtained from the fit is $\sim 10^{40}$ erg s$^{-1}$. From the above two luminosity values, the accretion efficiency ($\eta_{\rm acc}$) can be estimated to be 0.02. This value is low compared to 0.1 often used in the literature, which is unlikely to be the same for different systems.} However, the numerical simulations of ULX sources showed that the $\eta_{\rm acc}$ can be as low as 0.002 \citep{NarayanEtal2017MNRAS.469.2997N}. Therefore, a nearly IMBH accreting in super-Eddington regime can power a ULX at $\leq 10^{40}$ erg s$^{-1}$ when the accretion efficiency is low.

\subsection{Outflow properties and the ULX bubble}\label{sec:JetProp}
In this section, we use the model-fitted parameters to estimate different physical quantities of the mass outflows. 
The mass outflow to inflow ratio is estimated using the following relation \citep{Chak1999A&A...351..185C}, 

\begin{equation}
    \label{eq:outflowRate}
    R_{\dot m} = f_{\rm col}f_{o}^{3/2}\frac{R}{4}exp{\left(\frac{3}{2}-f_{o}\right)},
\end{equation}
where $f_{o}$ is $\frac{R^2}{R-1}$. 

\textcolor{black}{Our estimated $R_{\dot m}$ (in percent) for epochs A1 to A5 are 12.4$\pm$2.2, 15.6$\pm$1.8, 20.6$\pm$4.2, 12.0$\pm$2.3, and 18.1$\pm$3.0 respectively for the model fitted parameters $R$ and $f_{\rm col}$ in \autoref{table:JeTCAFResults}. In epoch A3, the higher outflow ratio and the smaller dynamic corona size explain that a significant amount of thermal energy has been taken away by the outflows, and the corona cooled down.} It can be possible that during this epoch the source was in the intermediate state as the shock compression ratio is in agreement with the theoretical range suggested in the model \citep[][]{Chak1999A&A...351..185C}. 

In addition to the above estimation and considering $\dot m_{\rm in}$ in the post-shock region or the corona, the jet/outflow rate is written as $R_{\dot m}$ $\dot m_{\rm in}$. Thereby, the jet/outflow power ($P_{\rm j}$) can be estimated using,

\begin{equation}
\label{eq:jet power}
P_{\rm j}= 1.3 \times 10^{38}\left(\frac{M_{\rm BH}}{M_{\odot}}\right) R_{\dot m}\; \dot m_{\rm in}\; \eta_j  ~~~~~\text{erg}~ \text{s}^{-1},  
\end{equation}

here, $\eta_j$ is the jet/outflow efficiency and M$_\odot$ is the mass of the Sun. \textcolor{black}{The values obtained for $P_{\rm j}$ across epochs A1 to A5 are (6.5$\pm$1.7, 5.0$\pm$1.0, 9.0$\pm$3.8, 6.5$\pm$1.8 and 7.8$\pm$1.8) $\eta_j \times$ 10$^{40}$ ergs s$^{-1}$ respectively. However, as we do not have $\eta_j$ beforehand, different values of it can give different $P_{\rm j}$. \citet[][]{gurpide2022arXiv220109333G} calculated the disc outflow power using nebula expansion rates and reported that the observed bubble has power $\sim$10$^{40}$ erg s$^{-1}$. Therefore, to compare with the observed estimation, $\eta_j$ has to be $\sim 0.1-0.2$.} 

\textcolor{black}{In addition, we have estimated the outflowing solid angle of $\sim 1.6\pi$ for the observed epochs using the inflow geometry and $f_{\rm col}$ parameter. A wide outflowing solid angle implies that the mass outflow is uncollimated which shaped the observed bubble.} We have further estimated the mass outflow velocity (${\rm v}_{\rm j}$), which varies as $\sqrt{T_{\rm shk}}$, as the shock is driving the outflow in \jetcaf\, model, where $T_{\rm shk}$ is the shock temperature (the proton temperature). The $T_{\rm shk}$ is estimated using the relation \citep{Debnathetal2014} $T_{\rm shk}= m_{\rm p} (R-1) c^{2}/2 R^{2} k_{\rm B} (X_{s}-1)$. Here $m_{\rm p}$ and $k_{\rm B}$ are the proton mass and Boltzmann constant respectively. \textcolor{black}{The calculated $T_{\rm shk}$ varies between $5.9-8.6\times 10^{10}$~K. Then equating the thermal energy with the kinetic energy of protons at the jet launching region, which is the CENBOL, ${\rm v}_{\rm j}$ comes out to be in between 0.1c - 0.2c. This is in accord with the results found by \citet[][]{walton2016ApJ...826L..26W}.} This velocity corresponds to absorption lines which originate from inner regions of the disc \citep[][]{walton2020MNRAS.494.6012W, Pinto2020MNRAS.492.4646P}. Therefore, our estimated mass outflowing angle and its velocity agree with previous observational results.

Using the above outflow quantities, we further estimate the age of the bubble ($t_{b}$) considering a free expansion of the shocked outflowing material \citep[][]{weaver1977ApJ...218..377W} through the ambient medium, which is given by
\begin{equation} \label{Eq:age}
t_{b} \simeq \left(\frac{\rho R_b^5}{\frac{1}{2}R_{\dot m}\dot m_{\rm in}{\rm v}_j^2} \right)^{1/3}.
\end{equation}
Assuming that the bubble is expanding through the neutral medium with a mean molecular weight $\mu=1.38$, and thus $\rho=\mu m_p n_{\rm ISM}$, where $n_{\rm ISM}$ is the hydrogen number density. The $R_b$ value of 134 pc and $n_{\rm ISM}=0.6$ cm$^{-3}$ are taken from \citet{gurpide2022arXiv220109333G}. Considering other jet quantities from the \jetcaf\, model fit (see \autoref{table:JeTCAFResults}), \autoref{Eq:age} gives the age of the bubble in the range \textcolor{black}{$\sim 3.3-6.5\times 10^5$ yr,} in agreement with the range suggested by \citet{gurpide2022arXiv220109333G}. We note that the mechanical power estimated in the denominator of \autoref{Eq:age} differs from the jet power estimated using \autoref{eq:jet power} as it estimates total power; both mechanical and thermal.

Hence we report that a nearly IMBH accreting at super-Eddington rates is able to explain the observational features and different time scales of formation and evolution of the ULX-1 bubble. ULX-1 is a suspected BH candidate as discussed in \autoref{sec:intro}. However, what has not been previously reported is the estimation of mass accretion by the central IMBH and the flow geometry using physical models. In principle, IMBH can accrete at super-Eddington rates. Though the existence of such IMBH is in dispute, and many proposed candidates are not widely accepted as definitive, these IMBHs might be necessary to explain the large gap in mass between StMBH and SMBHs. The strongest observational evidence for the existence of IMBHs was presented by \citet{FarrellEtal2009Natur.460...73F} in the edge-on spiral galaxy ESO\,243-49. Recently, studies through gravitational waves have reported BH mass of a 150 M$_{\odot}$ to exist \citep[][]{Gwaves2020PhRvL.125j1102A}. Some other studies also evidenced that SMBHs can accrete above their Eddington limit \citep[][and references therein]{PuDu2015ApJ...806...22D, hlieu2021ApJ...910..103L}. In \xmm\, spectral studies \citet[][]{jin2016MNRAS.455..691J} found the evidence of super-Eddington accretion onto RX\,J1140.1+0307, active galactic nuclei whose mass lies in the IMBH range ($\leq$ 10$^{6}$ M$_{\odot}$). 

\subsection{Limitations and Directions for Improvements}
The \jetcaf\,model fitted accretion parameters show that ULX$-$1 is a super-Eddington accretor harboring a nearly IMBH. Such super-Eddington accretion flows would lead to the formation of a strong wind perpendicular to the disc surface \citep{ShakuraSunyaev1973A&A....24..337S}. The radiation pressure in this accretion regime may drive the wind \citep{KingBegelman1999ApJ...519L.169K}, which can carry a large amount of mass from the disc. Likewise, the outflowing wind may also carry a significant amount of energy and angular momentum depending on the physical processes depositing them to the wind \citep[for radiatively inefficient flow,][]{BlandfordBegelman1999MNRAS.303L...1B}. 

Moreover, extracting information from the observations in the X-ray band is limited by our line of sight. Therefore, testing the degree of anisotropy of the X-ray emission remains challenging \citep[see][]{MiddletonEtal2021MNRAS.506.1045M}. A strong anisotropy is predicted in several theoretical studies in the super-Eddington accretion regime \citep{ShakuraSunyaev1973A&A....24..337S,poutenan2007MNRAS.377.1187P,NarayanEtal2017MNRAS.469.2997N}, which is still a poorly understood accretion regime. \textcolor{black}{Further, the present model does not include the disc inclination and spin parameter, that may affect the anisotropy effects, which are beyond the scope of the present work. Thus the current estimations of the model parameter values and the related physical quantities (\autoref{sec:JetProp}) may vary in detail modeling, keeping the parameter profiles unchanged. }

\section{Conclusion} 

We have conducted a joint \xmm+\nustar\, analysis of the well-known ULX NGC\,1313\,X-1 which evidenced BH at its center. We have used \jetcaf\, model to study the observed features of the accretion-ejection system. Our key findings are enlisted below:

\begin{itemize}
\item The mass accretion rates returned from the \jetcaf\, model fits to the data are super-Eddington, which is consistent with the earlier findings (\autoref{sec:intro}) that the ULX-1 is a super-Eddington accretor.

\item  \textcolor{black}{The mass outflow to inflow ratio estimated is $\sim 21\%$ - $12\%$ with the outflowing solid angle $\sim 1.6 \pi$.} Such a wide angle may indicate that the outflow is uncollimated which shaped the observed bubble, agrees with optical observations. 

\item \textcolor{black}{The possible BH mass returned from the data fitting is $133\pm33$ M$_\odot$, averaged over all observations. This implies that the ULX-1 harbors a nearly IMBH at its center. We redo the fitting by keeping the BH mass to 10, 30, and 100 M$_\odot$ and check the consistency of the goodness of the fit and the uncertainties of the model parameters. We find that the BH mass $>30 M_\odot$ returns a good fit, however, the uncertainty in model parameters improves at higher BH mass value.}

\item The super-Eddington accretion onto an IMBH can power a ULX at $\leq 10^{40}$ erg s$^{-1}$ if the accretion efficiency is low $\sim$ 0.02, however, their efficiency ($\sim 0.1-0.2$) increases when the jet/outflow is taken into account, consistent with numerical simulations in the literature \citep{NarayanEtal2017MNRAS.469.2997N}.

\item The \jetcaf\, model fitted parameters can explain the observed power of the recently discovered bubble \citep{gurpide2022arXiv220109333G} around the ULX-1, its age, and the wind launching velocity estimated from high-resolution spectroscopy \citep{Pinto2020MNRAS.492.4646P}.

\item According to possibilities discussed in \citet[][]{King2004MNRAS.347L..18K}, an IMBH can behave like a ULX when it accretes mass from any large mass reservoir with a high accretion rate $\gtrsim 10^{-6}$ M$_\odot$ yr$^{-1}$, consistent with our mass accretion rates. Thus, a fraction of ULXs discovered could be hosted by IMBHs.

The above conclusions are drawn from the X-ray spectral fitting using a physically motivated non-magnetic accretion-ejection based \jetcaf\, model. In the present model scenario, it emerges as a possibility of powering ULXs (or at least some ULXs) by super-Eddington accretion onto nearly IMBHs, which can also explain the ULX bubble properties. However, analysis of a large sample of ULXs is needed to further support this possibility. As discussed, some physical processes are required to implement in the modelling to further constrain the accretion-ejection parameters. 

\end{itemize}

\section*{Acknowledgements}
\textcolor{black}{We thank the referee for making constructive comments and suggestions that improved the quality of the manuscript.}
We gratefully acknowledge the Ramanujan Fellowship research grant (file \# RJF/2020/000113) by SERB, DST, Govt. of India for this work. This research has made use of the {\it NuSTAR} Data Analysis Software ({\sc nustardas}) jointly developed by the ASI Science Data Center (ASDC), Italy and the California Institute of Technology (Caltech), USA. This work is based
on observations obtained with {\it XMM-Newton}, an European Science Agency (ESA) science mission with instruments and contributions directly funded by ESA Member States and NASA. 
This research has made use of data obtained through the High Energy Astrophysics Science Archive Research Center Online Service, provided by NASA/Goddard Space Flight Center.

\section*{Data Availability}
Data used for this work is publicly available to NASA's HEASARC archive. \textcolor{black}{The \jetcaf\,model is currently not available in XSPEC, however, we are open to collaborate with the community. Presently, we are running the source code in XSPEC as a local model, it will be freely available in near future.}


\bibliography{ngc1313x1}{}
\bibliographystyle{aasjournal}



\end{document}